# A Brief Review of Ferroelectric Control of Magnetoresistance in Organic Spin Valves

Xiaoshan Xu*

*Department of Physics and Astronomy and Nebraska Center for Materials and Nanoscience, University of Nebraska, Lincoln, Nebraska 68588, USA*

## Abstract

Magnetoelectric coupling has been a trending research topic in both organic and inorganic materials and hybrids. The concept of controlling magnetism using an electric field is particularly appealing in energy efficient applications. In this spirit, ferroelectricity has been introduced to organic spin valves to manipulate the magneto transport, where the spin transport through the ferromagnet/organic spacer interfaces (spinterface) are under intensive study. The ferroelectric materials in the organic spin valves provide a knob to vary the interfacial energy alignment and the interfacial crystal structures, both are critical for the spin transport. In this review, we first go over the basic concepts of spin transport in organic spin valves. Then we introduce the recent efforts of controlling magnetoresistance of organic spin valves using ferroelectricity, where the ferroelectric material is either inserted as an interfacial layer or used as a spacer material. The realization of the ferroelectric control of magneto transport in organic spin valve, advances our understanding in the spin transport through the ferromagnet/organic interface and suggests more functionality of organic spintronic devices.

*xiaoshan.xu@unl.edu

# Introduction

Giant magnetoresistance has been a successful example of nanotechnology in which transport of spin-polarized current through interfaces is manipulated in nanoscale to vary the resistance of the devices. Wide application of this effect, such as in the read heads of the hard disks for much larger information density, has been realized; the fundamental research was awarded Nobel prizes in 2007. The concept of manipulating the spin degree of freedom of electrons to control the electrical transport, has now evolved into a large active field of research and technology, i.e. spintronics, with application potentials in information storage and processing, sensors, energy generation, etc.[1–5]

The effect of giant magnetoresistance can manifest in a trilayer junction shown in Fig. 1(a). The junction contains a non-magnetic (NM) spacer layer sandwiched by two ferromagnetic (FM) electrodes. Depending on the alignment of the magnetization of the two FM electrodes, the resistance of the junction changes between high and low values. By controlling the alignment of the magnetization of the two FM electrodes, the resistance of the junction can be switched by an external magnetic field. The trilayer junction can then be regarded as a spin valve, in the sense that the electrical current can be turned on and off using the external magnetic field. Depending on the spacer material and thickness, three categories of spin valves are mostly studied. 1) If the spacer material is a metal, the magnetoresistance (MR) is expected to be small compared with the volume resistance[6]. Superlattice-fashioned structures were adopted to enhance MR by increasing the number of interfaces while keeping the thickness of the junction and the volume resistance constant[7–11]. In this case, the thickness of the spacer is often less than a few nanometers; the MR has to do with the indirect exchange coupling between the magnetic layers[12,13]. 2) If the spacer is a thick semiconductor, the two FM layers are magnetically decoupled and the transport through the spacer becomes diffusive. In this case, the MR hinges on spin injection, which is actually difficult for a metal/semiconductor ohmic contact. Tunneling through a barrier between the FM and the semiconductor provides a more efficient route for the spin injection[14–17]. 3) If the spacer is a thin insulator, the electrical transport is based on spin-conserved tunneling between the two electrodes. Therefore, the MR is related to the alignment of spin polarization of the initial and final states of the tunneling.

Organic spin valves are trilayer structures including organic semiconductors or insulators as the spacers materials. The long spin life time of the organic materials[18,19] (due to the weak spin-orbit coupling in the light elements such as carbon and hydrogen), is appealing for spin transport. In addition, the flexibility, environment friendliness, and the vast chemical diversity of organic materials suggest great application potentials of organic spintronic devices. Organic spin valves generally belong to the latter two categories introduced above, where the two FM electrodes are decoupled in terms of exchange interactions[6]. The alignment of the magnetization of the two FM electrodes, can be tuned by an external magnetic field, based on their difference in magnetic coercivity. The MR has a butterfly-like shape, as illustrated in Fig. 1(b) and (c). If the resistance of the spin valve is high (low) when the magnetization of the two FM layers are antiparallel, it is called normal (inverse) or positive (negative) MR.

Encouraged by early promising results on organic spin valves[20,21], efforts on understanding the fundamental mechanism and on realizing organic spintronic devices, has been growing rapidly. However, several key issues, such as spin injection and spin polarization at the FM/organic interfaces, are still not fully understood in organic spin valves. To tackle these key issues, being able to tune the crystal structure and electronic structure at the FM/organic interfaces

(spinterface[22]) appears to be critical, because the sign and magnitude of the MR is determined by the spin polarization at these interfaces, which are sensitive to the electronic structure. The electronic structure at the FM/spacer interface is a result of coupling between that of the two materials, which is based on the crystal structure and the energy level alignment of the two materials. Ferroelectric (FE) materials, either as an interfacial material or as the spacer itself, offer tunabilities to the electronic structures at the interface, by changing the energy level alignment or the interfacial crystal structure. Besides the important role in studying fundamental mechanism, the FE controlled organic spin valves also have potential applications in multi-state information storage devices and energy efficient information processing, because the changes caused by the FE material, via switching the electrical polarization using an electric field, are non-volatile.

Here we review the FE control of MR in organic spin valves as one of the frontiers of organic spintronics. Inspired by similar work in inorganic spin valves and previous efforts in tuning FM/organic interfaces[23–31], recently, a sequence of work has studied the multistates in the organic spin valves using FE materials as the spacer or as an interfacial layer. [32–35] The results confirm the critical role of energy alignment and interfacial structure in the spinterface. We first introduce the concepts and fundamental processes in the diffusive and tunneling spin valves, followed by the specific problems in organic spin valves. To review the ferroelectric control of MR in organic spin valves, we divide the effects of FE materials on the interface into electrostatic change of energy alignment and the interfacial structural change due to the polarization reversal.

## Diffusive spin valves

A spin valve can be realized in a junction where electrical current flows diffusively, which means that the electrochemical potential of the charge carriers changes when they travel through the junction. To understand the MR of the diffusive spin valves, it is better to discuss the boundary resistance at the FM/NM interface first. In the two-current (spin up and spin down) model[36], the fundamental difference between the FM and the NM, are in the equilibrium spin polarization ($\alpha$), defined as the proportion of the spin-up branch of the current, which represents the proportion of spin-up density of states at the Fermi level [Fig. 1(d)]. For the FM and NM layers, one has $\alpha_F \neq 0.5$ and $\alpha_N = 0.5$ respectively. When the spin polarized current flows from the FM layer into the NM layer, it tends to depolarize toward the equilibrium state of the NM layer. As shown in Fig. 2(a), the dynamics of the spin polarization can be modelled using the difference of electrochemical potentials of the spin-up and spin-down branch of the current, assuming no interfacial spin scattering. If we assume $\alpha_F > 0.5$ (spin up carriers are the majority), near the FM/NM interface, the electrochemical potential of the spin-up branch has to be higher than that of the spin-down branch, to drive the population from the former to the latter. This is true also on the NM side of the interface [Fig. 2(a)], indicating that the charge carriers are also spin-polarized near the interface on the NM side [Fig. 2(b)]. The spin polarization in the NM layer and its generation via the electrical transport through the FM/NM interface are referred to as spin accumulation and spin injection respectively[1,5].

The mean electrochemical potential of the current is discontinuous at the interface [Fig. 2(a)], which is the origin of the boundary resistance ($R_B$), given as

$$\frac{1}{R_B} = \frac{1}{(\alpha_F - \alpha_N)^2}\left[\frac{\alpha_F(1-\alpha_F)}{\lambda_F/\sigma_F} + \frac{\alpha_N(1-\alpha_N)}{\lambda_N/\sigma_N}\right],$$

where $\lambda_F$ and $\lambda_N$ are the spin diffusion length[37,38] of the FM and NM layers respectively, $\sigma_F$ and $\sigma_N$ are the conductivity of the FM and NM layers respectively. The boundary resistance can also be understood in terms of the population change between the two current branches, or virtual currents. In general, the virtual currents run on both the FM side and the NM side of the interface in parallel over the distance about the spin diffusion length, as shown in Fig. 2(b). The additional resistance ($R_B$) of the interface due to the virtual current (change of spin polarization) is then the two resistances $R_{NB} = \frac{\lambda_N}{\sigma_N} \frac{(\alpha_F - \alpha_N)^2}{\alpha_N(1-\alpha_N)}$ and $R_{FB} = \frac{\lambda_F}{\sigma_F} \frac{(\alpha_F - \alpha_N)^2}{\alpha_F(1-\alpha_F)}$ in parallel. Notice that if $\alpha_F = 1$ (half metal FM), only $R_{NB}$ contributes to the boundary resistance. Accordingly, the change of spin polarization of the charge carriers occurs only in the NM layer and the spin injection is 100%.

In a FM/NM/FM trilayer structure, if the NM layer is thin enough compared with $\lambda_N$, the spin polarization at the two FM/NM interfaces, interferes[39]. In the case that the spin polarizations of the two FM layers are along the same direction [Fig. 2(c) and 2(d)], the current may not have to totally depolarize at one interface before it is polarized at the other interface. The difference between the electrochemical potentials of the two current branches is then smaller; the discontinuity of the mean electrochemical potential and the boundary resistance, which both come from the change of spin polarization, are reduced compared with those when the spin polarizations of the two FM layers are along the opposite directions [Fig. 2(e) and 2(f)]. If the thickness of the NM layer $d$ is much smaller than $\lambda_N$ ($\frac{d}{\lambda_N} \ll 1$) and the two FM materials are identical, one has $R_B^{\uparrow\uparrow} = \frac{\alpha_{F1}(\alpha_{F1} - \alpha_N)}{2\alpha_N(1-\alpha_N)} \frac{\lambda_N}{\sigma_N} \frac{d}{\lambda_N}$ and $R_B^{\uparrow\downarrow} = \frac{(2\alpha_{F1} - 1)^2}{2\alpha_{F1}(1-\alpha_{F1})} \frac{\lambda_{F1}}{\sigma_{F1}}$ for the parallel and antiparallel alignment respectively; hence $R_B^{\uparrow\downarrow} \gg R_B^{\uparrow\uparrow}$. This change of boundary resistance due to the change of relative alignment of the spin polarization of the two FM layers, is the mechanism of the giant magnetoresistance in the classical two-current model.

Notice that it is alignment between the spin polarizations of the charge carriers of the two FM layers instead of that between the magnetization of the two FM layers, that determines the MR. The MR sign of a spin valve can be analyzed in terms of quantum mechanical description of the spin polarization, defined as $P = \frac{n_M - n_m}{n_M + n_m}$, where $n_M$ and $n_m$ are the number of states of the majority spin and minority spin respectively of the states of certain enrgy. For typical metallic ferromagnets like Co, at the Fermi level, most states have the minority spin, corresponding to $P < 0$[40]. For the oxide conductor La$_{0.7}$Sr$_{0.3}$MnO$_3$ (LSMO), which is often used as an FM electrode, the states at Fermi level corresponds to $P > 0$[41]. To analyze the MR sign, one defines the interfacial spin polarization $P_1^*$ for the carrier injection as the $P$ of the initial states. Similarly, the interfacial spin polarization of the carrier collection $P_2^*$ is the $P$ of the final states. The MR sign can be predicted from $P_1^* P_2^*$, where positive (negative) values of the product correspond to normal (inverse) MR. If the spacer is a non-magnetic metal, interfacial spin corresponds to the spin polarization at the Fermi level of the FM materials. If the spacer is a semiconductor, the interfacial spin polarization depends on the detailed coupling between the electronic structures of the FM and spacer material.

The MR of the trilayer diffusive spin valve comes from the change of boundary resistance. The absolute value of the boundary resistance is on the order of the resistance of the junction materials (two FM layers and one NM layer) over the length scale of the spin diffusion length, which may be small compared with the volume resistance in the all-metal junctions. Hence, the MR is not expected to be very large in the all-metal trilayer diffusive spin valves. Unfortunately, the MR cannot be enhanced simply by replacing the spacer with materials of much larger resistivity. According to Eq. (1), the total boundary resistance is the resistance of the FM and NM channels in parallel. So, increasing only the NM resistance does not change the total boundary resistance significantly. In fact, the large NM resistance reduces the spin injection into the NM layer, i.e. the depolarization of the spin current occurs mostly in the FM layer because the FM channel has much smaller resistance. Hence, the two FM/NM interfaces are decoupled and little MR is expected. This is the famous impedance-mismatch problem[42].

## Tunneling spin valves

The impedance-mismatch problem is absent in the tunneling spin valves. When the NM layer is insulating and thin enough, the mechanism of electrical conduction becomes quantum tunneling, in which the spin states of the carriers are conserved. The transition from initial states in FM1 to final states in FM2, depends on the relative spin polarization of these states. Therefore, one may observe resistance change when the relative alignment of the magnetization of the two FM electrodes is manipulated by the external field.

Similar to that in the diffusive spin valve, it is the spin polarization of the initial and final states that matters for the MR sign. Since the tunneling resistance is much larger than the resistance of the electrodes, the estimation of MR may ignore the volume-resistance contribution from the electrodes. Hence, in the Julliere's model[43], MR is predicted as $\frac{2P_1P_2}{1-P_1P_2}$, where $P_1$ and $P_2$ are the spin polarization of the initial and final states of the tunneling in the two FM electrodes. According to this model, much larger MR is possible compared with that in the diffusive spin valves. Indeed, the tunneling MR up to 600% has been reported by Ikeda et al, in CoFeB/MgO/CoFeB spin valves at 300 K[44].

Under bias voltage, the initial and final states in FM1 and FM2 may shift with respect to the Fermi level. Since the spin polarization of states at different energy levels may be different, the MR may change with the bias voltage. De Teresa et al studied LSMO/STO/Co tunneling spin valve[40], where STO stands for $SrTiO_3$. At low bias voltage, the negative $P$ in Co and the positive $P$ in LSMO results in an inverse MR. At higher bias, different part of the states of the Co participate in the tunneling, significantly different MR, both in magnitude and in sign, was observed. The bias-voltage dependence of the MR reflects the spin density of states of Co.

The electronic structure of the tunneling barrier (spacer) also plays an important role in the tunneling MR. In the case of Co, the spin polarization $P$ of the $s$ and $d$ states at the Fermi level are positive and negative respectively[40,45], while in LSMO, the Fermi level is only occupied by the $d$ states, which has a positive spin polarization $P$[41]. If the barrier is $Al_2O_3$, the tunneling of $s$ electrons from and to Co Fermi level is favored, correspond to a positive MR. If the barrier is STO, the $d$ carriers in Co is more involved in the tunneling, causing a negative MR[45].

## Organic spin valves

In an organic spin valve, the spacer is an organic insulator or semiconductor. One can construct organic tunneling spin valve using a thin organic spacers[46–51]. For example, Barraud et al reported a 300% MR in LSMO/Alq$_3$/Co tunneling spin valve at 2 K[52], where Alq$_3$ stands for tris-(8-hydroxyquinoline) aluminum; Santos et al reported an 8% tunneling MR in a similar junction at room temperature[46], both in tunneling organic spin valves.

If the organic layer is too thick for the charge carriers to tunnel, diffusive conduction through the junction is expected. Since the organic insulator and semiconductors all have much larger resistance than that of the electrodes, the impedance mismatch problem discussed above could minimize the MR. Surprisingly, a sizable MR (~10%) have been repeatedly observed in the organic spin valves using LSMO as one of the FM electrodes[20,53–60]. To resolve the controversy, spin injection, i.e., whether the charge carriers in the organic spacer is spin polarized, has been intensively tested.

Widely accepted demonstration of spin injection employs the Hanle effect[61–66]. To show the Hanle effect in a trilayer spin valve, a magnetic field perpendicular to the magnetization of the FM electrodes is applied. The spin of the charge carriers is expected to precess due to the magnetic field when they travel through the spacer. The period of the precession is inversely proportional to the external magnetic field. Therefore, the spin polarization of the charge carriers after going through the spacer can be tuned using the magnetic field; the resistance of the junction oscillates with respect to the transverse magnetic field[67–70]. Hanle effect has been successfully employed to demonstrate spin injection into inorganic semiconductors using the hot electron injection[67–71]. On the other hand, it has not been observed in organic spin valves[72], despite that fact that the spin injection has been demonstrated by muon spin resonance[73,74] and two-photon photoemission[75].

Although the issue certainly has not been settled, several possibilities may reconcile the observed spin injection and apparent impedance mismatch. One possibility is that, in the LSMO/Alq$_3$/Co junctions, the spin injection can be greatly enhanced due to the high spin polarization of charge carriers in LSMO (half metal)[20,21,41,53–58,76,77]. Indeed, the MR of the NiFe/Alq$_3$/FeCo junction is much smaller than that in the LSMO/Alq$_3$/Co junctions[73,78,79]. Another possibility is that the injection of spin polarized current into the organic spacer comes from tunneling through a barrier between the FM electrode and the spacer before the current transports diffusively in the spacer[14,15][80].

Compared with inorganic materials, the spin transport through organic spin valves is more incoherence[81–83], which may affect the measurement of Hanle effect. Nevertheless, the Hanle effect has been observed in non-local structure of inorganic junctions, where the spin transport is totally incoherent[62]. Therefore, it was speculated that the absence of Hanle effect is due to a time scale of spin transport that is smaller than that of the spin precession[72]. Following this logic, if one assumes that the spin and charge transport in the same speed, a unreasonable mobility for the charge transport is needed.[72] Therefore, model with decoupled charge and spin transport has been proposed, in which the spin transport relies on exchange interactions.[84,85] Experimental evidence of exchange-mediate spin transport was later reported.[86]

## Ferroelectric interfaces

Ferroelectric (FE) materials are crystalline materials that exhibit spontaneous electrical polarizations switchable by an external electric field[87]. The electric polarization originates from the displacement of positive and negative charge centers, which is allowed only when the crystal structure lacks inversion symmetry. Therefore, the switching of polarizations of ferroelectric materials involves both displacement of atoms and the corresponding shift of charges; both can be employed in active control of interfacial properties.

The existence of dipole, i.e. separated positive and negative charges, generates uneven electric potential in space. As illustrated in Fig. 3(a), in an FE film where polarization is pointing perpendicular to the film plane, the electric potential undergoes a rapid change across the film due to the electric field inside the material. The surface potentials are $\phi_{S\pm} = \pm \frac{\sigma_P}{2\varepsilon\varepsilon_0} d$, where $\sigma_P$ is the surface charge of the FE, $\varepsilon_0$ and $\varepsilon$ are the vacuum dielectric constant and the relative dielectric constant of the FE respectively, and $d$ is the thickness of the FE layer[88]. Outside the FE film, the potential decays with a length scale much larger than the film thickness. When another material is in contact with an FE material, its charge distribution and the electrochemical potentials at the interface will be affected. For example, at an FE/metal interface, the charge in the metal accumulates to screen the charge of the FE and cancel the electric field. Fig. 4(b) illustrates the charge distribution of a metal/FE/metal junction. Using the Thomas-Fermi model of screening and assuming the two metals are electrically shorted, one obtains the electric potentials on the two metal sides that decay exponentially: $\phi_1 = \frac{\sigma_s \delta_1}{\varepsilon_0} e^{\frac{x}{\delta_1}}$ and $\phi_2 = -\frac{\sigma_s \delta_2}{\varepsilon_0} e^{-\frac{x-d}{\delta_2}}$, where $\delta_1$ and $\delta_2$ are the screening lengths in the metal 1 (M1) and metal 2 (M2) respectively, and $\sigma_s = \frac{d\sigma_P}{\varepsilon(\delta_1+\delta_2)+d}$ is the screening charge[25,88]. When the screening length of the metal is much larger than the thickness of the FE, one has $\sigma_s \ll \sigma_P$ and the potentials at the FE/metal interfaces are basically the same as that of the FE surface potential. When the screening length is much smaller than that of the thickness of FE, one has $\sigma_s \approx \sigma_P$ and the potential at the FE/metal is greatly reduces. For "good" metallic materials like Co and Fe, the screening length is less than an angstrom, which suggests a complete screening of charge, or zero potential at the interface. The screening at the two FM/metal interfaces can be different and cause asymmetric electric potentials. As shown in Fig. 4(d), the larger screening length results in smaller screening charge and larger interfacial potential.

The electric potential in Fig. 4(c) determines the vacuum potential of electrons in the junction materials. The uneven vacuum potential causes the bending and shift of energy bands in the materials in contact with the FE [Fig. 4(d)], similar to that in the Schottky contact between a metal and a semiconductor[89,90]. In the case of Schottky contact, the accumulation of charge at the interface is the response to the chemical potential (Fermi energy) difference[89,90], while it is the response to the electrostatic potential difference that causes the charge accumulation at the FE/metal interface. Although it is common for the interface of two materials of different work functions to have charge accumulation and interfacial dipoles[91,92], the FE materials enables the tunability of the electrochemical potentials at the interface using an external electric field in a nonvolatile fashion.

## Control of spin transport using ferroelectric interfaces
### Diffusive organic spin valves and interfacial energy alignment
In a diffusive spin valves using an organic semiconductor (OSC) spacer, the energy alignment at the FM/NM interfaces plays an important role in the spin transport, by affecting the interfacial spin polarization $P^*$. First, the alignment between the Fermi level of the metals and the molecular energy levels of the OSC, determines the charge carrier type in the electric transport through the junction. As shown in Fig. 5(a), if the metal Fermi energy is closer to the lowest unoccupied molecular orbital (LUMO), the main charge carriers are electrons. If the metal Fermi energy is closer to the highest occupied molecular orbital (HOMO), the main charge carriers are holes. Second, in the carrier collection process, energy level of the final states in the metal is determined by the energy level of carriers in the OSC. For an FM material, because the spin polarization of states at different energy levels may be different, the spin polarization of carrier collection $P^*$ is likely different for different carrier types and different energy alignment between the OSC and the FM.

The alignment between the metal Fermi energy and the HOMO/LUMO of OSC can be adjusted simply by using metals of different work functions as the electrodes. This effect has been studied by comparing the magneto transport of the Co/Alq$_3$/NiFe junction and the Co/Ca/Alq$_3$/Ca/NiFe junction[26]. For Fe, Co, and Ni, the work functions are large (5.0 eV[93]) and the Fermi energy is closer to the HOMO of Alq$_3$. So, the hole transport is expected in the Co/Alq$_3$/NiFe junction. The work function Ca is much smaller (2.9 eV[93]) and the Fermi energy is closer to the LUMO of Alq$_3$. So, the transport of electrons through LUMO of Alq$_3$ is expected. Experiments show significantly smaller resistance in the Co/Ca/Alq$_3$/Ca/NiFe junction. In addition, a much larger MR with a reversed sign in Co/Ca/Alq$_3$/Ca/NiFe was observed, indicating change of charge carrier type between these two junctions.

The energy alignment between the metal and the OSC can also be manipulated by inserting a dipolar layer between the metal electrode and the OSC. Using LiF as the dipolar materials, Schultz et al studied the magneto transport in the NiFe/Alq$_3$/FeCo and the NiFe/LiF/Alq$_3$/FeCo junctions[73]. The spin polarization in the Alq$_3$ was measured using muon spin resonance. It was found that, by inserting the LiF layer (1 nm), the spin polarization of the carrier collection from Alq$_3$ to NiFe electrode changes from negative to positive. Consequentially, the sign of the MR is also reversed: while the NiFe/Alq$_3$/FeCo junction shows a negative MR, the NiFe/LiF/Alq$_3$/FeCo shows a positive MR. These observations are consistent with the existence of a large interfacial dipole due to the insertion of LiF layer. Since Alq$_3$ is a semiconductor with a screening length much longer than the spacer thickness, the vacuum potential generated by the LiF dipole is maintained and shifts the HOMO/LUMO of Alq$_3$ at the interface. In the NiFe/Alq$_3$/FeCo junction, the collected holes from Alq$_3$ HOMO goes into an energy level with mostly spin minority states ($P^*<0$). In contrast, due to the shift of HOMO downwards by the LiF layer, the collected holes from Alq$_3$ HOMO may go into an energy level with mostly spin majority states ($P^*>0$). This change of spin polarization of carrier collection leads to the change of MR sign.

If the dipole inserted between the FM electrode and the OSC comes from an FE material, the effect on the magneto transport is tunable by an electric field, due to the switchability of the FE dipole. This effect has been studied by Sun et al in LSMO/PZT/Alq$_3$/Co junctions[34], where PZT stands for PbZr$_{0.2}$Ti$_{0.8}$O$_3$, an oxide FE material with a large electric polarization (~80 µC/cm$^{-2}$)[94]. It was found that the MR depends on the initial of voltage applied on the junction. The bias dependence of the MR, the $MR(V_{MR})$ function, was used to gauge the effect of the initial voltage, where $V_{MR}$ is the measurement voltage of the MR. The results show that by applying a positive (negative) initial voltage $V_{MAX}$ that is larger than the measurement voltage $V_{MR}$, the $MR(V_{MR})$ function of the LSMO/PZT/Alq$_3$/Co junction with as-grown PZT, shifts toward the negative (positive) voltage directions [Fig. 5 (a)-(d)]; the sign of MR is always negative. The shift become larger when the initial voltage $V_{MAX}$ is larger. More importantly, this shift of the $MR(V_{MR})$ function could not be observed in several control junctions, including LSMO/Alq$_3$/Co, LSMO/STO/Alq$_3$/Co, and LSMO/PZT/Co. These observations are consistent with the presence of an interfacial dipole due to the PZT layer between LSMO and Alq$_3$. Since Alq$_3$ is a semiconductor with poor screening ability, the dipole from PZT is maintained and causes a shift of vacuum potential (Δ) of Alq$_3$; the effective voltage applied on Alq$_3$ becomes $V_{eff} = V_{MR} + \Delta$. The initial voltage $V_{MAX}$ changes the dipole and shifts the vacuum potential. Positive (negative) $V_{MAX}$ makes

Δ more positive (negative), corresponding to the shift of $MR(V_{MR})$ function toward the negative (positive) $V_{MR}$ direction. It is understandable that the shift of the $MR(V_{MR})$ function is not observed in the LSMO/Alq3/Co and LSMO/STO/Alq3/Co junctions, because there is no interfacial FE dipole. For the LSMO/PZT/Co junction, the initial voltage does change the MR, but there is still no shift of the $MR(V_{MR})$ function, because the dipole of PZT is expected to be screened by the metal electrodes. The change of MR in the LSMO/PZT/Co junction is more likely caused by the change of electronic structure at the interface due to the atomic displacement of the FE rather than the shift of vacuum potentials.

Dramatic changes were observed in the magneto transport of the LSMO/PZT/Alq3/Co junction, after a larger electric field that switches the FE polarization of PZT was applied. First, the sign of $MR(V_{MR})$ function changes from all negative in the LSMO/PZT/Alq3/Co junction with as grown PZT (polarization pointing up toward Alq3) [Fig. 6(a)], to mostly positive after the FE polarization is switched to pointing down toward LSMO [Fig. 6(c)]. Second, the effect of $V_{MAX}$ is not a simple shift of the $MR(V_{MR})$ function; instead, it may change the sign of the MR. The dependence of the MR sign on $V_{MR}$ and $V_{MAX}$ is in line with varying the polarization of PZT in minor polarization-voltage loop. It appears that, when the polarization is lower than a certain value, the MR becomes positive, as illustrated in Fig. 6(b). This dependence of MR sign on the polarization of PZT can be understood using the change of energy alignment between LSMO and Alq3 due to the dipole moment of PZT. As shown in Fig. 6(d), due to the poor screening ability of Alq3, the electric dipole moment of PZT generates a shift of the vacuum potential. When the PZT polarization is pointing up, the injection and collection of holes at the LSMO/PZT/Alq3 interface is between the spin majority band of LSMO and Alq3 HOMO. In contrast, when the PZT polarization is pointing down, the HOMO of Alq3 is shifted up. The higher energy states of LSMO with minority spin becomes accessible for the injection and collection of holes at the LSMO/PZT/Alq3 interface. Therefore, the shift of energy states in A1q3 by the PZT dipole, may change the sign of interfacial spin polarization $P^*$, which changes the sign of the MR of the LSMO/PZT/Alq3 junction.

Tunneling organic spin valves and the interfacial crystal structure
In a FM/FE/FM tunneling spin valve, since the surface charge of the FE is most likely screened by the FM metal, the effect of the FE on the vacuum potential and energy alignment is minimized. Besides changing the charge distribution, the switching of FE dipole is also accompanied by the atomic displacements. As mentioned above, in the tunneling spin valve, the coupling between the barrier and the FM electrode in electronic structures may affect the interfacial spin polarization significantly. Therefore, by changing the coupling of electronic structures at the FM/FE interface, the atomic displacement may change the interfacial spin polarization and the MR.

This effect has been investigated by Liang et al in LSMO/PVDF/Co junctions[33], where PDVF stands for poly(vinylidene fluoride)[95]. The ferroelectricity of PVDF originates from the long-range order of the dipoles in the polymer and the switching corresponds to the collective rotation of dipoles along the molecular chains. As shown in Fig. 7(a), when the electric polarization of PVDF is switched to pointing down toward LSMO, the MR is positive. In contrast, when the polarization is pointing up toward Co, the MR is negative. In addition to MR, the resistance of the junction also changes significantly due to the reversal of polarization of PVDF. As shown in in Fig. 7(b), the relation between the initial voltage and the resistance measured at a low voltage (10 mV), is plotted. A clear hysteresis is observed; the change of resistance (ER) due to the initial voltage is up to 75%. At the same time, the MR also shows a hysteretic behavior with respect to the initial voltage [Fig. 7(c)]. It appears that the direction of the electric polarization is correlated with the sign of MR, although the magnitude of the MR is larger when the electric polarization is pointing up toward Co. The asymmetry in the MR is also observed in the bias voltage dependence [Fig. 7(d)] and temperature dependence [Fig. 7(e)].

The dependence of MR on the initial voltage and the asymmetry in magnitude is explained in terms of the change of electronic structure at the PVDF/Co interface caused by the reversal of electric polarization of PVDF. As mentioned above, the reversal of electric polarization of PVDF corresponds to the rotation of the molecular dipole along the polymer chain. When the electric polarization is pointing toward Co, it is the hydrogen (H) atoms that are in direct contact with the Co. Otherwise, when the electric polarization is pointing toward LSMO, it is the fluorine (F) atoms that are at the Co/PVDF interface. Therefore, the structure of the PVDF/Co interface can be switched between H-C-F/Co and F-C-H/Co correspond to the FE polarization pointing to LSMO and Co respectively [Fig. 8(a) and (b)]. The density of states of PVDF at the PVDF/Co interface has been calculated. Since these states are gap states induced by the interface with Co, whether F or H atom are in direct contact with Co makes a significant difference. It was found that at the H-C-F/Co interface, although the spin polarization of the gap states of the first layer of PVDF is negative (the same sign as that in Co), the second layer gap states show a positive spin polarization. For the F-C-H/Co interface, both the first and second layer in PVDF show negative spin polarization for the gap states. These results explain the reversal of MR sign when the electric polarization of PVDF is reversed. In addition, it was found that the F-C-H/Co interface is more energetically stable than the H-C-F/Co interface, which is consistent with the observation that MR vanishes at lower temperature when the polarization is pointing toward LSMO (H-C-F/Co interface) than that when the polarization is pointing toward Co (F-C-H/Co interface) [Fig. 7(e)].

The effect of FE polarization on tunneling FM/FE/FM junction was also studied in LSMO/ P(VDF-TrFE)/Co junctions at 200 K, where P(VDF-TrFE) stands for ferroelectric copolymers poly(vinylindene fluoride-trifluoroethylene)[32]. As shown in Fig. 9(a)-(c), the tunneling resistance of the junction depends on both the magnetic field and the initial pulsed electric field. The effect of the electric field increases with the magnitude of the field, consistent with the behavior of effect of electric field on the electric polarization. The effect of the electric field on MR also increases with the field, as shown in Fig. 9(d). Although the MR sign does not change when the electric polarization of the FE reverses, the asymmetry agrees with the previous results in the LSMO/PVDF/Co junctions[33]. Similar to that in the LSMO/PVDF/Co junctions, when the electric polarization is pointing up toward Co, a negative MR with larger magnitude is observed. When the electric polarization is pointing down toward LSMO, the sign of MR remains negative but the magnitude decreases significantly. These results are consistent with the modification of interfacial state of the FE spacer due to the different coupling between the FE and FM electrode when the electric polarization is switched.

## Conclusion and outlook

The recent efforts of exploiting ferroelectric control of MR in organic spin valves show encouraging indication of effects on the energy alignment and on the interfacial crystal structures. However, since the exact mechanism of spin transport through organic spin valves is still not fully understood, more work on similar devices with other ferroelectric materials[95–97] and more characterization on the interfacial crystal structures and on the interfacial electronic structures are needed to elucidate the ferroelectric effect as well as other concomitant effects such as the change of magnetism of the electrodes[98–100] and change of transport mechanism[101–103]. At the same time, as the field of organic spintronics keeps growing, the spin polarization at the FM/organic interfaces (spinterface) will remain a focus. More examples of tuning the spinterface by varying the energy level alignment or by varying the interfacial crystal structure are expected to be investigated. These investigations will be not only important for organic spintronics; they will also be useful for tuning other electronic devices (such as light emission diode and photovoltaics) involving organic/inorganic interfaces.

## Acknowledgements
This project was primarily supported by the National Science Foundation through the Nebraska Materials Research Science and Engineering Center (Grant No. DMR-1420645).

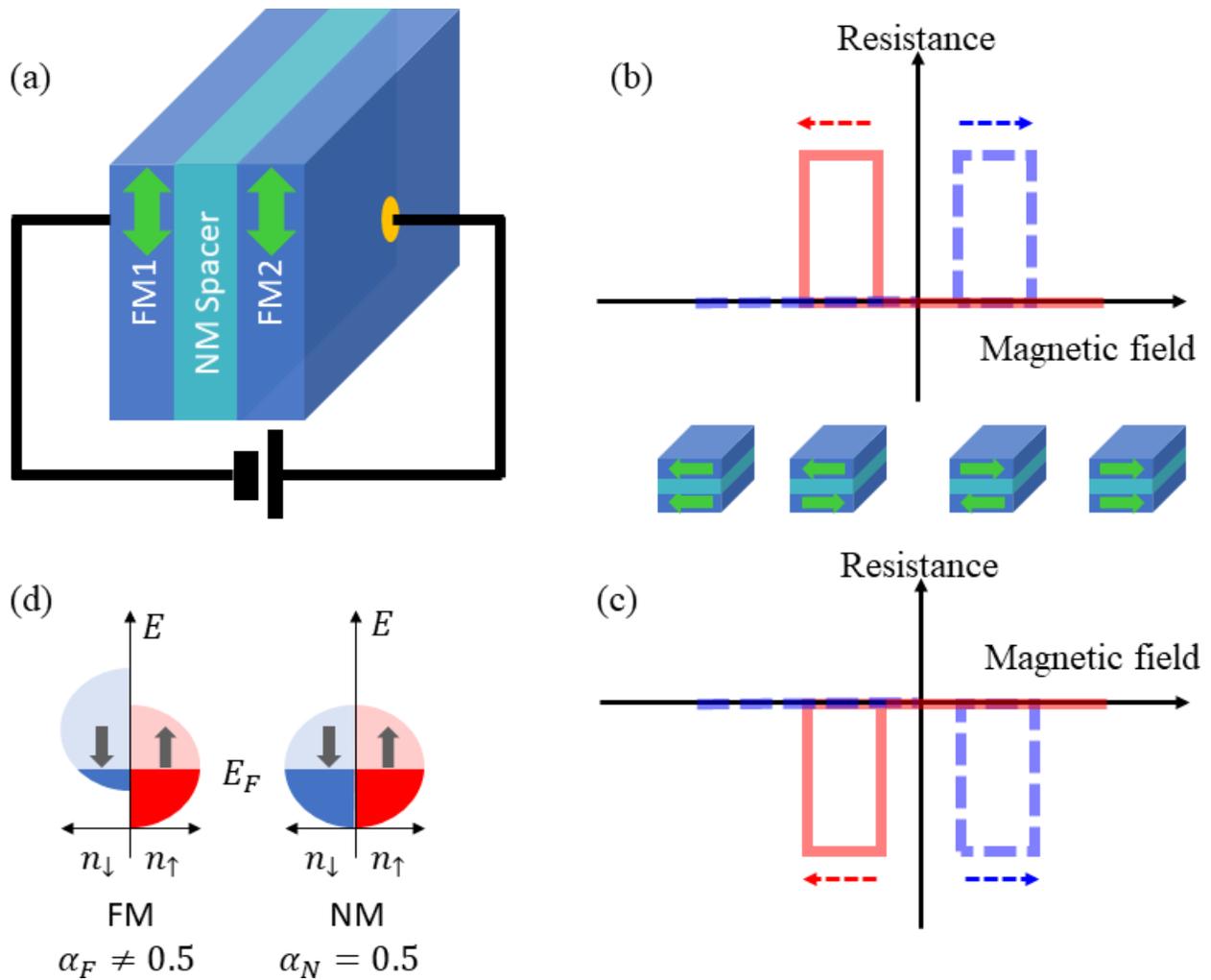

**Figure 1**. (a) Schematics of an FM/NM/FM trilayer spin valve. (b) and (c) are the magnetic-field dependence of resistance of the spin valve for positive (normal) and negative (inverse) MR respectively. (d) Schematics of the electronic structure of the FM and NM materials, where $n_\uparrow$ and $n_\downarrow$ are the number of states of the up spin and down spin respectively.

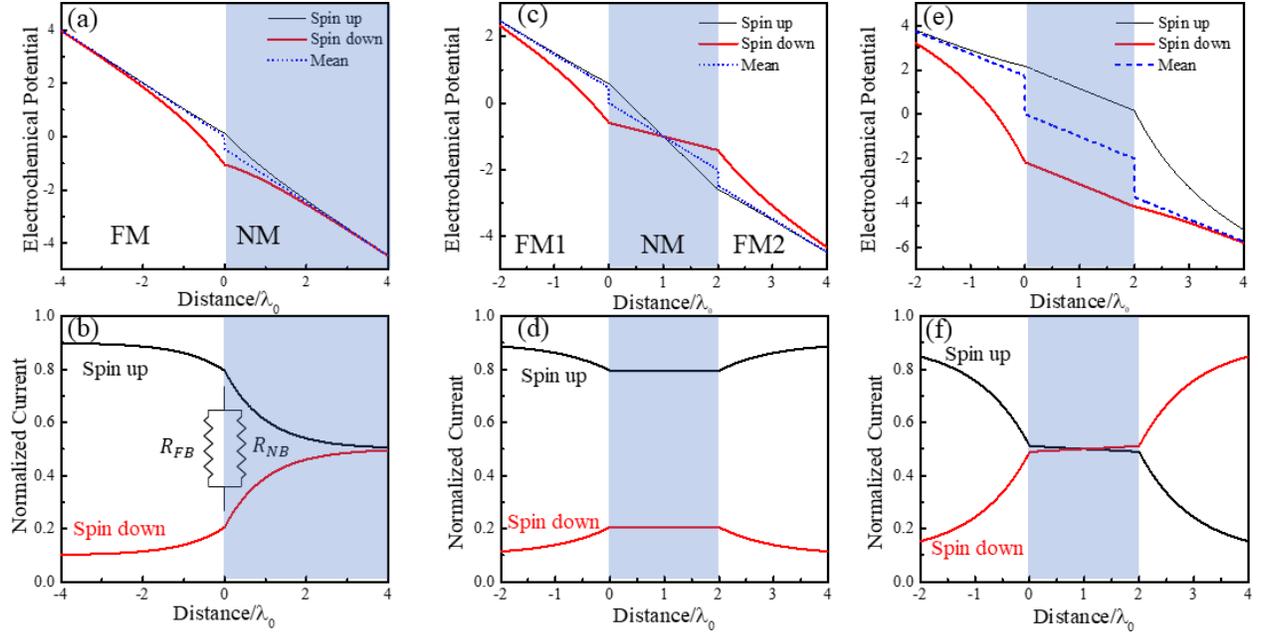

**Figure 2**. Calculated spatial dependence of the electrochemical potential and the normalized current in the two-current model. (a) and (b) are near a FM/NM interface. (c) and (d) are for a trilayer spin valve when the spin polarization of the two FM layers are aligned ($\alpha_{FM1} > 0.5, \alpha_{FM2} > 0.5$). (e) and (f) are for a trilayer spin valve when the spin polarization of the two FM layers are anti-aligned ($\alpha_{FM1} > 0.5, \alpha_{FM2} < 0.5$). The calculation assumes: $\lambda_{F1} = \lambda_{F2} = \lambda_0$, $\lambda_N \gg \lambda_0$, and $d = 2\lambda_0$, where $\lambda_0$ is a length scale.

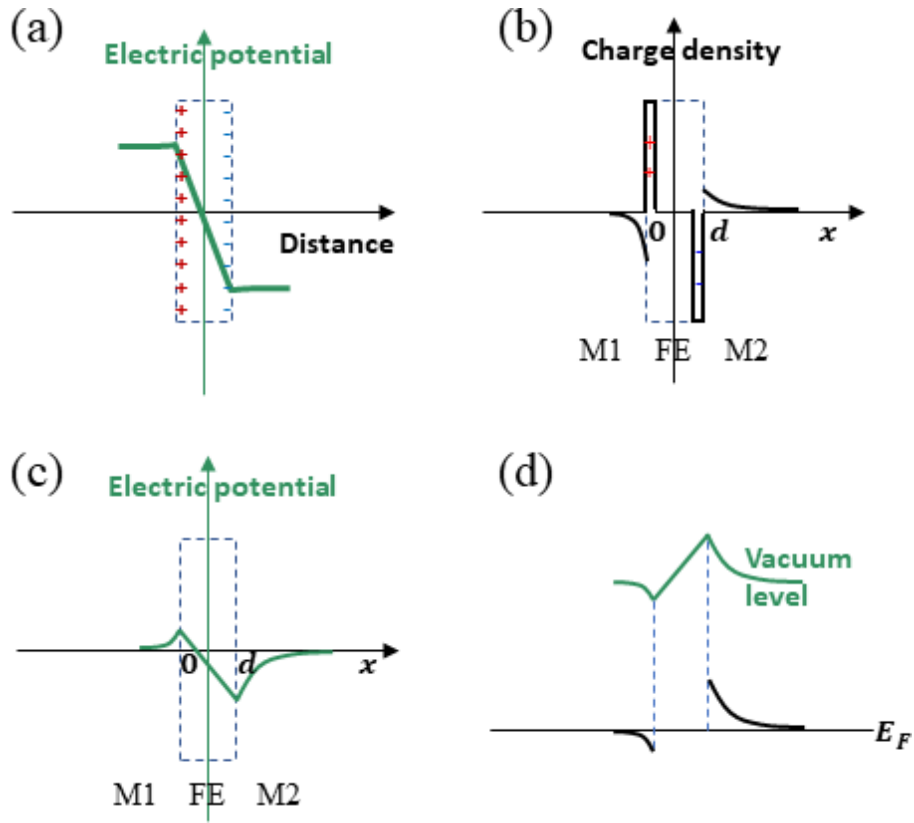

**Figure 3**. (a) Spatial dependence of the electric potential caused by a ferroelectric film (indicated using the charge distribution). (b) and (c) are the charge distribution and the vacuum electric potential of a metal/FE/metal junction. (d) Energy level bending and shift at the metal/FE/metal junction. The lower curve represents the metal states that are at the Fermi energy far away from the FE.

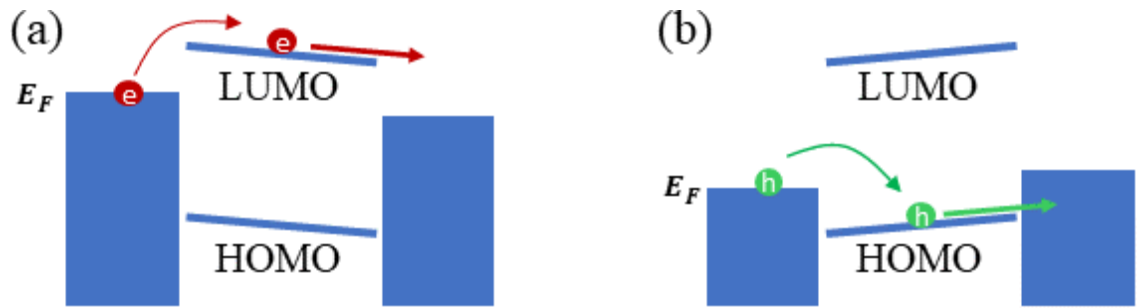

**Figure 4**. The schematics of the charge transport across a metal/OSC/metal junction. (a) When the Fermi energy of the metal is close to the LUMO of OSC, electron transport is favored. (b) When the Fermi energy of the metal is close to the HOMO of OSC, hole transport is favored.

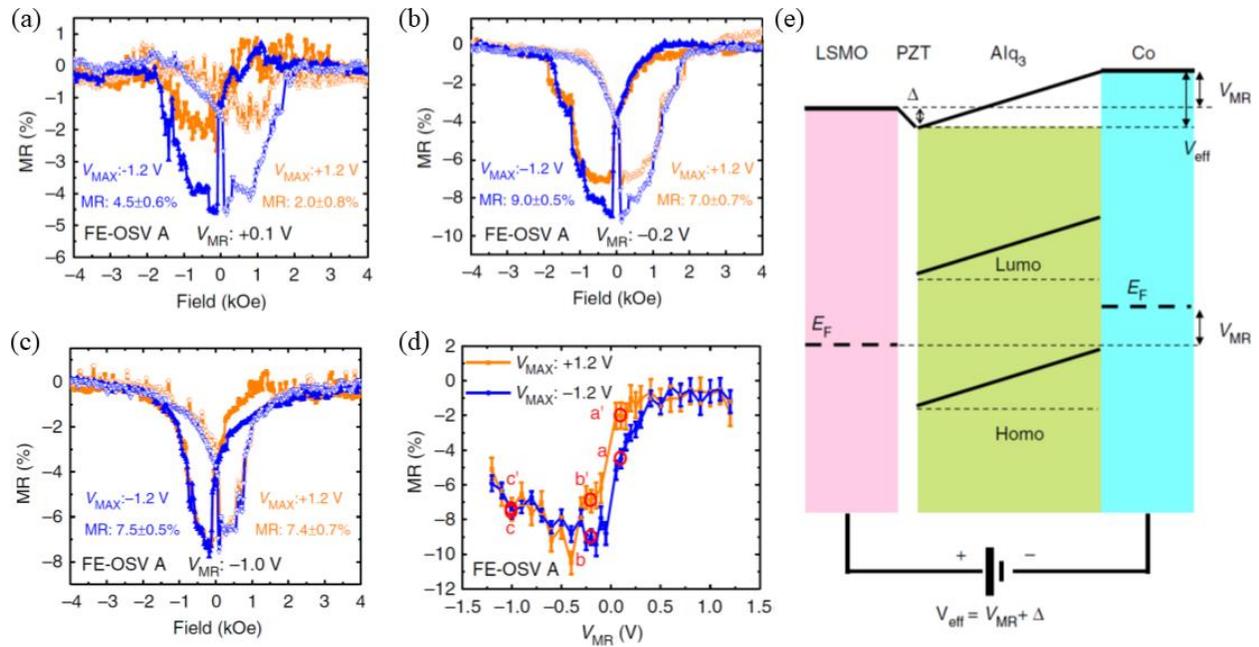

**Figure 5**. MR of the LSMO/PZT/Alq$_3$/Co junction. (a)-(c) are the MR measured at 0.1, -0.2, and -1.0 V respectively when a 1.2 or a -1.2 V initial voltage is applied. (d) is the MR as a function of measurement voltage ($V_{MR}$) when a 1.2 or a -1.2 V initial voltage is applied. (e) is the schematic diagram indicating that the shift of vacuum level and the change of effective voltage on the OSC caused by the FE dipole of PZT.

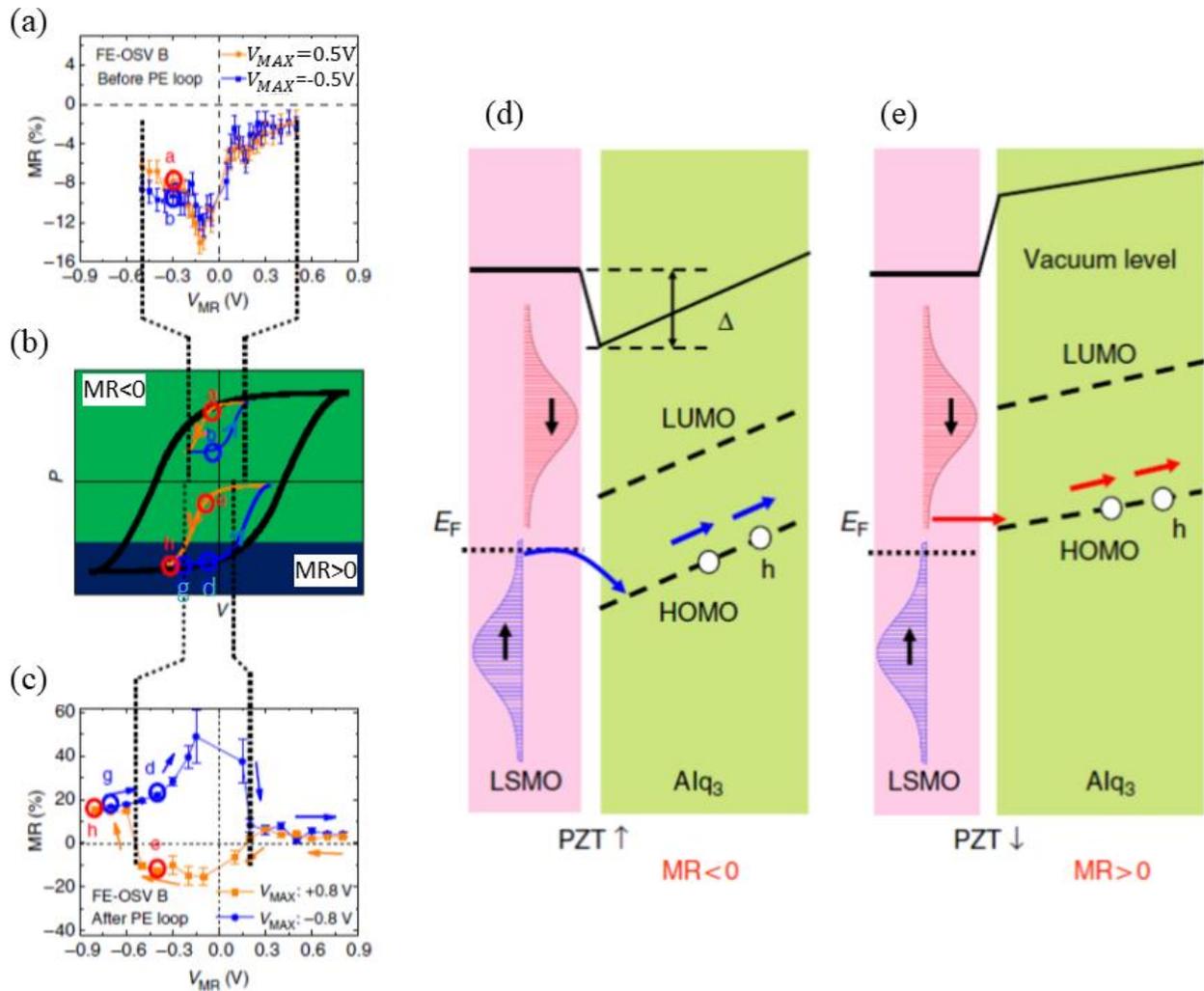

**Figure 6**. (a) MR as a function of measurement voltage ($V_{MR}$) when a 0.5 or a -0.5 V initial voltage is applied. (b) Schematics of the polarization-voltage hysteresis loop of PZT and the minor loops. (c) MR as a function of measurement voltage ($V_{MR}$) when a 0.8 or a -0.8 V initial voltage is applied. Before the measurement, the PZT was poled to negative saturation polarization. (d) and (e) are the energy level alignment of the LSMO/PZT/Alq$_3$ interface when PZT polarization is pointing toward Alq$_3$ and toward LSMO respectively.

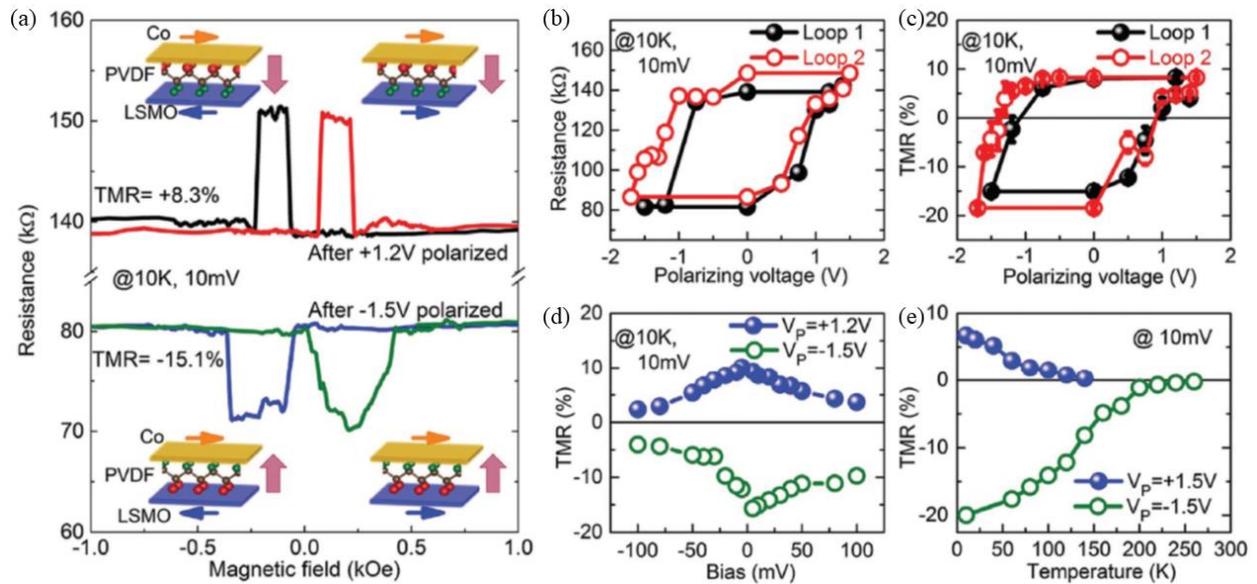

**Figure 7**. MR of the LSMO/PVDF/Co junction. (a) MR measured when the PVDF polarization is pointing toward Co and toward LSMO respectively. (b) Resistance of the LSMO/PVDF/Co junction as a function of the poling voltage. (c) Tunneling MR as a function of the poling voltage. (d) Tunneling MR as a function of measurement voltage for 1.2 and -1.5 V poling voltages respectively. (e) Tunneling MR as a function of temperature for 1.5 and -1.5 V poling voltages respectively. Reproduced with permission from Ref. [33]. Copyright 2016, Wiely.

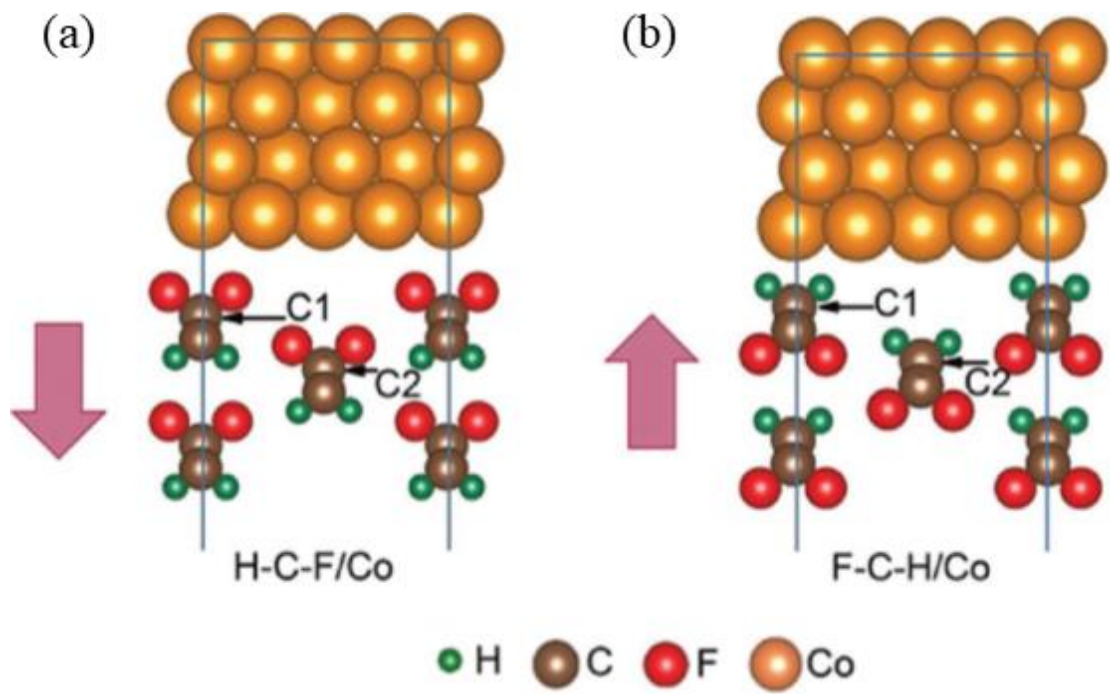

**Figure 8**. Crystal structure at the Co/PVDF interface when the polarization is pointing away (a) and toward (b) Co respectively. Reproduced with permission from Ref. [33]. Copyright 2016, Wiley.

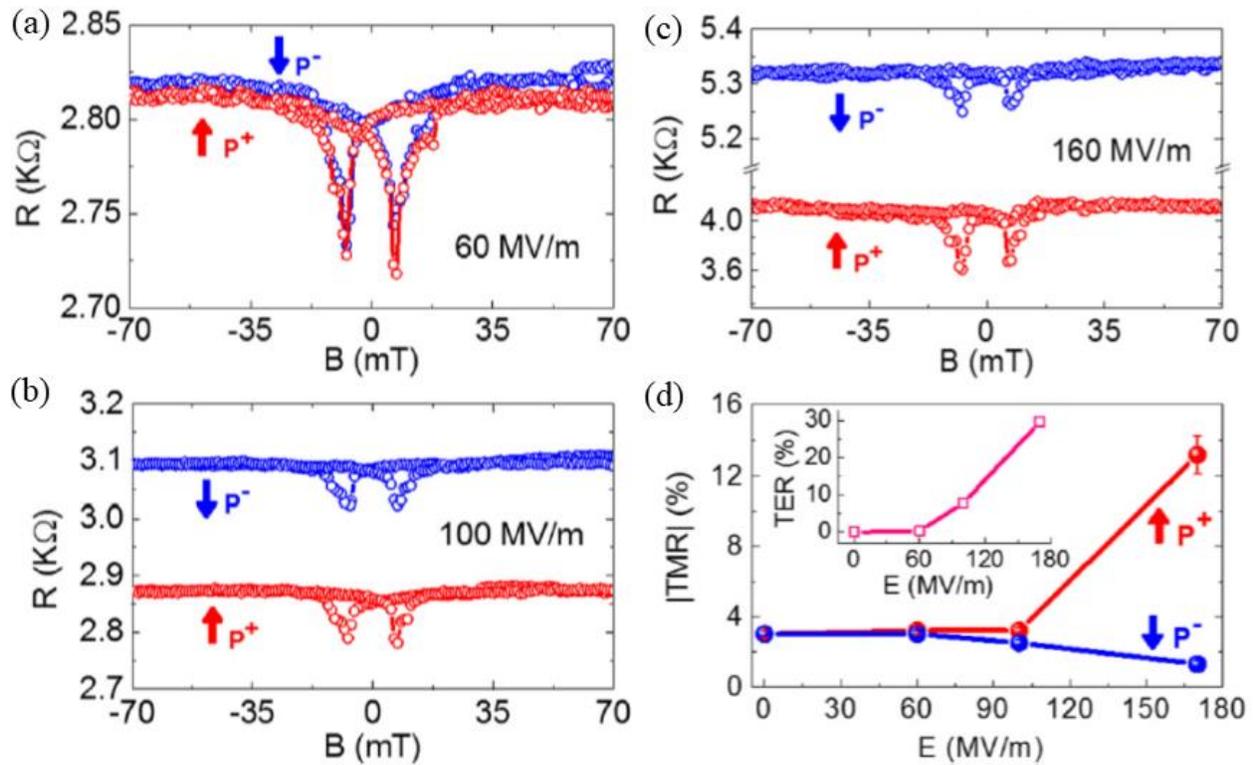

**Figure 9**. MR of the LSMO/P(VDF-TrFE)/Co junction. (a)-(c) MR for positively and negatively poled P(VDF-TrFE) using 60, 100, and 160 MV/m respectively. (d) Tunneling MR and tunneling electroresistance (ER) as a function of poling voltage for poling fields. Reproduced with permission from Ref. [32]. Copyright 2017, the American Institute of Physics.